\documentstyle[prb,aps,multicol,epsf]{revtex}

\begin{document}
\draft
\newcommand {\be}{\begin{equation}}
\newcommand {\ee}{\end{equation}}
\newcommand {\bea}{\begin{eqnarray}}
\newcommand {\eea}{\end{eqnarray}}
\newcommand {\nn}{\nonumber}

\newcommand{\boxed}[1]{\fbox{$\displaystyle{#1}$}}
\newcommand{\Hi}{\mbox{${\cal H}$\/}}
\newcommand{\Cr}{\mbox{$A^*$}}
\newcommand{\Spur}{\mbox{Spur}}
\newcommand{\bra}[1]{\mbox{$ \langle #1 | $}}
\newcommand{\ket}[1]{\mbox{$ | #1 \rangle $}}
\newcommand{\braket}[2]{\mbox{$ \langle #1 | #2 \rangle $}}
\newcommand{\expect}[1]{\mbox{$ \langle #1 \rangle $}}
\newcommand{\An}{\mbox{$A$}}
\newcommand{\Va}{\mbox{$\Omega$\/}}
\newcommand{\Real}{\mbox{$I{\!}R$\/}}
\newcommand{\Complex}{\mbox{$/{\!\!}C$\/}}
\newcommand{\Poincare}{\mbox{$P_+^\uparrow$\/}}
\newcommand{\Wi}{\mbox{${\cal W}$\/}}
\newcommand{\Schwartz}{\mbox{${\cal S}\/$}}
\newcommand{\Fock}{\mbox{${\cal F}\/$}}
\newcommand{\srcacuo}
        {\mbox{$\mbox{Sr}_{14-x}\mbox{Ca}_{x}\mbox{Cu}_{24}\mbox{O}_{41}$}} 
\newcommand{\srtcacuo}
        {\mbox{$\mbox{Sr}_{3}\mbox{Ca}_{11}\mbox{Cu}_{24}\mbox{O}_{41}$}}
\newcommand
{\sracuo}{\mbox{$\mbox{Sr}_{14-x}\mbox{A}_{x}\mbox{Cu}_{24}\mbox{O}_{41}$,
        (A=Ca, Y)}}
\newcommand
{\cuoplane}{\mbox{$\mbox{Cu}_{2}\mbox{O}_{3}$}}
\newcommand{\srcuo}{\mbox{$\mbox{Sr}\mbox{Cu}_{2}\mbox{O}_{3}$ }}
\newcommand{\srqcuo}{\mbox{Sr$_{14}$Cu$_{24}$O$_{41}$ }}
\newcommand{\bcd}{\tilde{c}^{\dag}}
\newcommand{\bc}{\tilde{c}}
\newcommand{\ba}{{\mathbf a}}	
\newcommand{\ve}{{\mathbf e}}	
\newcommand{\bi}{{\mathbf i}}
\newcommand{\bj}{{\mathbf j}}
\newcommand{\bk}{{\mathbf k}}
\newcommand{\bl}{{\mathbf l}}
\newcommand{\br}{{\mathbf r}}
\newcommand{\bR}{{\mathbf R}}
\newcommand{\bS}{{\mathbf S}}
\newcommand{\bx}{{\mathbf x}}
\newcommand{\kb}[1]{b_{#1}^{\dag}}
\newcommand{\bb}[1]{b_{#1}}
\newcommand{\bv}{{\mathbf v}}
\newcommand{\hc}{{\mathrm h.c.}}
\newcommand{\Jperp}{J_{\perp}}
\newcommand{\Jpar}{J_{\|}}
\newcommand{\gs}{\ge}
%
%

\title{Electronic Structure of Ladder Cuprates}

\author{ T.~F.~A.~M\"uller$^{(1)}$, V.~Anisimov$^{(1)}$, T.~M.~Rice$^{(1)}$,
I.~Dasgupta$^{(2)}$, and T.~Saha-Dasgupta$^{(2)}$}

\address{$^{(1)}$ Institut f\"ur Theoretische Physik,
ETH-H\"onggerberg, CH-8093 Zurich, Switzerland \\
$^{(2)}$ Max--Planck--Institut f\"ur Festk\"orperforschung, D-70569 Stuttgart,
Federal Republic of Germany } 

\date{\today}
\maketitle
\begin{abstract}
We  study the electronic structure of the ladder compounds 
\srcacuo\ and \srcuo. LDA calculations for both give similar
 Cu 3d-bands near the Fermi energy. The
 hopping parameters 
estimated by fitting  LDA energy bands
show a strong anisotropy between the $t_{\perp}$ and $t_{\|}$ intra-ladder
hopping and small inter-ladder hopping.
A  downfolding method shows
 that this anisotropy arises from 
 the ladder structure. 
The  conductivity perpendicular to the ladders is computed assuming
incoherent tunneling giving a  value
close to experiment. 
\end{abstract}

\preprint{ETH-TH/98-03}
\begin{multicols}{2}
\narrowtext

\begin{figure}
	\centerline{\epsfxsize=8.cm\epsfbox{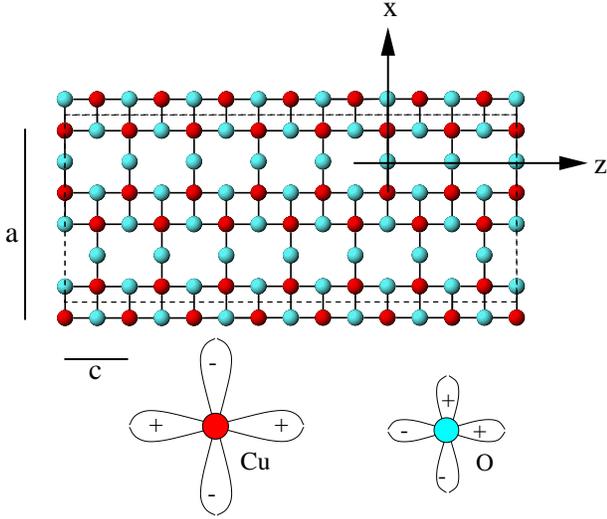}}
\caption{The Cu$_2$O$_3$ ladder plane. } 
\label{ladderp}
\end{figure}

 \srcacuo\ \cite{mccarron,siegrist} is the first material in which  doped ladder 
(see Fig.~\ref{ladderp})
can be experimentally studied and compared to theoretical predictions
of a Luther-Emery state with a spin-gap, hole-pairing
and superconductivity \cite{dagotto,rice,gopa,tsunetsugu,Troyer,HaywardPoil}. 
A spin-gap of $\Delta \simeq 280$K (for $x=9$)
has been measured\cite{magishi} and superconductivity 
under high pressure $P > 3$GPa has been
found \cite{magishi,uehara} in Ca-rich samples $x \simeq 11$(Ca11),
having ladder layers doped with $20\%$ holes ($\delta\simeq0.2$)
\cite{magishi,osafune,mizuno}.
	The transport properties are  dominated by holes in the
ladder  
planes. The normal state of the Ca11
 shows a strong anisotropy 
between the dc-resistivity along and across the ladder direction with
$\rho_{\perp}/\rho_{\|}\simeq 30 $ at T=100K~\cite{motoyama}. 
	For lower temperature, both resistivities increase exponentially 
due to localization effects. 
	For higher temperatures, $\rho_{\|}$ increases linearly in $T$
while $\rho_{\perp}$ remains nearly constant,
$\rho_{\perp}=12\mathrm{m}\Omega \mathrm{cm}$. 
	 The mean free
path along the ladder is larger than the lattice constant while across the
ladders it is smaller than the inter-ladder distance indicating incoherent
transport in this direction.

	Moreover, fits of the spin susceptibility have shown a large
difference between the  exchange coupling $\Jperp(\Jpar)$
along the rungs(legs) of the
ladder \cite{Johnston,Troyerpc} even if both involve similar $180^0$ Cu-O-Cu
superexchange processes. 
Analysis of neutron scattering data gives \mbox{$\Jperp=72$meV} and
\mbox{$\Jpar=130$meV} \cite{eccleston}. 
	 For these reasons a detailed
examination of the electronic structure is desirable.

In this paper we present LDA calculations of the electronic structure
which give  estimates of effective hopping matrix elements
between states on different copper ions.
	The LDA studies are performed for SrCu$_2$O$_3$, a
compound which possesses the same kind of Cu$_2$O$_3$ ladder planes as  
\srcacuo (see Fig.~\ref{ladderp}).
Recently, Arai and Tsunetsugu \cite{Arai} reported LDA calculations
for M$_{14}$Cu$_{24}$O$_{41}$,
(M=Sr or Ca) which give similar results.

\begin{figure}
\centerline{\epsfxsize=14cm\epsfbox{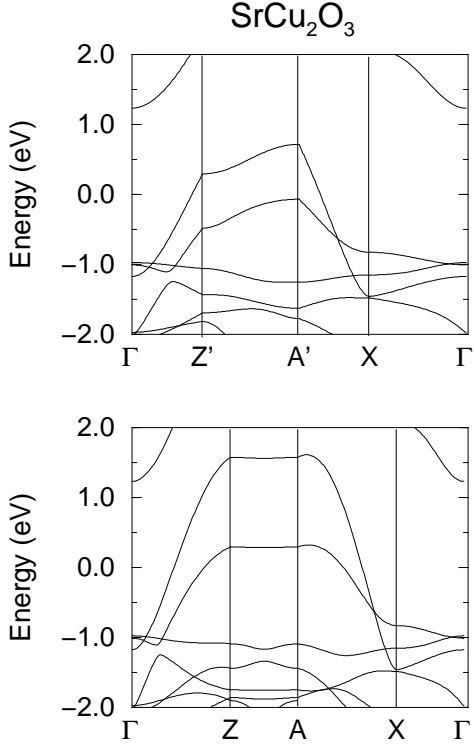}}
\caption{LDA band calculations on the path \mbox{$\Gamma Z'A'X \Gamma$}
(uppermost graph) 
and the path \mbox{$\Gamma Z A X \Gamma$} (lower graph).}
\label{LDA}
\end{figure}

The TB-LMTO ASA \cite{andersen0} energy bands for \srcuo
are plotted in Fig.~\ref{LDA}.
The uppermost graph displays the bands on the path
$\Gamma=(0,0,0)$ $Z'=(0,0,\pi/2c)$, $A'=(2\pi/a, 0,\pi/2c)$, and
$X=(2\pi/a,0,0)$. 
The two parallel bands near zero energy (Fermi energy of the
half-filled band)
are separated from the rest of the spectrum. 
They are due to hybridization through $\sigma$-bonds
of the $2p$ O- and  $3d_{x^2-z^2}$ Cu-orbitals. 
These bands hybridize with non-bonding O-bands near $\Gamma$.
In the lower graph, the energy bands are shown up to the edge of
the zone with \mbox{$Z=(0,0,\pi/c)$} and
\mbox{$A=(2\pi/a,0,\pi/c)$}. Due to destructive interference,
the 2 D-bands do not display dispersion on the path $ZA$.

	The low energy physics can be described by an
effective model containing only  these two bands with similar shape
near the Fermi energy. Such a model includes  
 only one state per Cu with 
effective hopping matrix elements. These 2 bands are the bonding (b) 
and anti-bonding (a) rung bands of the effective ladder model.

Note
that the parallel nature of the bands at  $k_z=\pi/2c$ (path Z'A')
cannot be explained by 
effective interactions between nearest-neighbor (n.n.) Cu-sites only. In such
a model, 
the  dispersion along the $k_x$ direction is given by the inter-chain hopping
matrix element between the second leg of one ladder and the first leg of 
the next ladder.  
Since the hopping matrix elements between two bonding(antibonding) states
of the rung $r$  of two neighboring ladders $l$ and $l'$
$b_{\sigma,l,r}(a_{\sigma,l,r})=\frac{1}{\sqrt{2}}(\phi_{\sigma,l,r,1}+(-)
\phi_{\sigma,l,r,2})$ 
is given by 
\begin{eqnarray}
	\expect{b_{\sigma,l,r}H(t_{ll'})b_{\sigma,l',r}}& = &
	\expect{\phi_{\sigma,l,r,2}H(t_{ll'})\phi_{\sigma,l',r,1}},\nonumber\\
	\expect{a_{\sigma,l,r}H(t_{ll'})a_{\sigma,l',r}}& = &
	-\expect{\phi_{\sigma,l,r,2}H(t_{ll'})\phi_{\sigma,l',r,1}},
\label{equ:1}
\end{eqnarray}
where the last index of $\phi_{\sigma,l',r,1}$ labels the ladder 
leg. Therefore b- and a-states should have  an opposite dispersion in the
$k_x$-direction.  
Thus, an effective atomic model must contain some
longer range inter-ladder hopping to account for their parallel nature.

Arai and Tsunetsugu introduced a  simpler rung parameterization of the band structure fitting the b- 
and a-bands separately, allowing n.n.\ and n.n.n.\ hopping leading
to the forms\cite{newkx}, 
\begin{eqnarray}
	\epsilon(\bk)& = &\epsilon_0-2h_{\|,1}\cos(k_z)-2h_{\|,2}\cos(2k_z)-
	\nonumber 
	\\
	& & [4h_{\perp,1}\cos(\frac{1}{2}k_z)+4h_{\perp,2}\cos(\frac{3}{2}k_z)]
	\cos(k_x).\label{araieq}
\end{eqnarray}
The values they obtained for \srqcuo are in good qualitative agreement
with ours for \srcuo (see Table~\ref{tabArai}).
Note that the signs of the interladder hopping parameter $h_{\perp,1}$ does not
change between an a-band and b-band contrary to the expectations from 
Eq.~\ref{equ:1}. 
All hopping parameters apart from $h_{\|,1}$ are higher order in the Cu-O
($t_{pd}$) and O-O overlaps $t_{pp}$ and thus much smaller.

\begin{figure}
\centerline{\epsfxsize=8.5cm\epsfbox{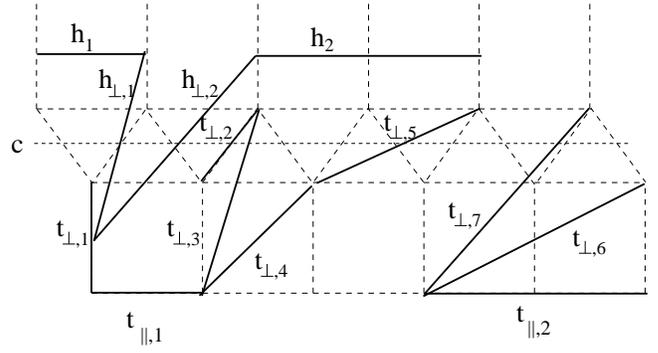}}
\vspace{0.5cm}
\caption{Coupled ladders illustrating effective hoppings between Cu sites (t)
or Cu-Cu rungs (h)}
\label{hopping}
\end{figure}

 To gain more insight we  introduce a single
parameterization of both bands in terms of intersite
hopping parameters shown in Fig.~\ref{hopping}
The solution of this tight-binding model is
\begin{eqnarray}
	\lefteqn{\epsilon_{\pm}(\bk)  =  \epsilon_0 +
	\epsilon_{\|}(k_z)+\epsilon_{\perp,1}(k_z)\cos(k_x)}  
	 \label{eq:tight} \\\nonumber
	& & \pm \sqrt{\epsilon_{\perp,3}(k_z)^2+\epsilon_{\perp,4}(k_z)^2+
	2\epsilon_{\perp,3}(k_z)\epsilon_{\perp,4}(k_z)\cos(k_x)}
\end{eqnarray}
where
\begin{eqnarray}
	\epsilon_{\|}(k_z) & = &
	-2t_{\|,1}\cos(k_z)-2t_{\|,2}\cos(2k_z),\nonumber \\
	\epsilon_{\perp,1}(k_z)& =&
	-4t_{\perp,3}\cos(\frac{k_z}{2})
	-4t_{\perp,7}\cos(\frac{3k_z}{2}),\nonumber \\
	\epsilon_{\perp,3}(k_z) &= & 
	t_{\perp,1}+2t_{\perp,4}\cos(k_z)+2t_{\perp,6}\cos(2k_z), \nonumber \\
	\epsilon_{\perp,4}(k_z) &= & 
	2t_{\perp,2}\cos(\frac{k_z}{2})+2t_{\perp,5}\cos(\frac{3k_z}{2}). 
\label{equ:2}
\end{eqnarray}
The parallel nature  of the bands at $k_z=\pi/2$ is recovered if
the term in the square root of Eq.~\ref{eq:tight} is independent of $k_x$. 
This  obtains if  $\epsilon_{\perp,4}(\pi/2)\simeq 0$, 
 leading to $t_{\perp,2}, t_{\perp,5} \simeq 0 $ or to 
$t_{\perp,2}=t_{\perp,5}$. In these cases the dispersion 
at $k_z=\pi/2$ is due to the $\epsilon_{\perp,1}$ term being a function of 
$t_{\perp,3}$ and $t_{\perp,7}$. 
It clearly shows that an intersite model must
contain at least  the third order hopping term $t_{\perp,3}$. 

	The coupled ladder system has a glide symmetry  given by 
the product of a reflection through the $c$-axis (see Fig.~\ref{hopping})
and a translation of half a lattice constant along the ladder.
When this operation is applied twice it is equivalent
to a translation of one lattice constant along the ladder.
	This implies that the energy band at $k_x=0$ and $k_x=\pi$
can be parameterized through one parameter. Actually 
$\epsilon_{\pm}(k_x=\pi,k_z)=\epsilon_{\pm}(k_x=0,2\pi-k_z)$ as can be directly
checked from Eq.~\ref{eq:tight}. This  allows one to use a single function
containing the information about all hopping parameters.
  This symmetry  also implies the lack of dispersion in the
$k_x$ direction at $k_z=\pi$ as discussed above, as well as the square root
form  containing intra- 
and inter-ladder hopping terms. 

Introducing $\sigma=1(-1)$ for $k_x=0(\pi)$
Eq.~\ref{eq:tight} reduces to the simpler form
\begin{eqnarray}
	\epsilon_{\pm,\sigma}(k_z) & = & \epsilon_0 +
	 	\epsilon_{\|}(k_z)+\sigma \epsilon_{\perp,1}(k_z) \nonumber \\
	& & \pm (\epsilon_{\perp,3}(k_z) +\sigma \epsilon_{\perp,4}(k_z)),
	\label{simeq}
\end{eqnarray}
representing the 4 energy bands of the double-ladder system.
Rewriting the bands from the rung form (\ref{araieq}) to the intersite form 
(\ref{simeq}) gives the values of Table~\ref{par}.
They are consistent with each other
and emphasize  the 
dominance of the n.n.\ intra-ladder matrix elements $t_{\|}$ and
$t_{\perp}$ w.r.t.\ the others. 
Moreover, they show surprisingly that these two parameters describing n.n.\ 
Cu-Cu processes differ from each other by
$\sim 35\%$.

Recently, Andersen {\it et al.}\cite{andersen1,andersen}
 introduced a systematic downfolding
scheme to obtain an effective single (or few) band model capable of
reproducing  the details of the LDA bands close to the Fermi level.
Effective hopping parameters are
calculated by performing the Fourier transform of the downfolded Hamiltonian 
$H(\bk)\longrightarrow H(\bR)$,
for $ |\bR|$ less than a cut-off radius $R_0$.
This has the advantage among others that
it allows the origin of the parameters in the effective single or few band 
 model to
be traced. We have applied this scheme to the LDA bands for \srcuo. 
The anisotropy $t_{\|,1}\neq t_{\perp,1}$ is best understood by starting
with an effective model 
including the $3d_{x^2-z^2}$, $4s$ Cu- and $2p_x$, $2p_z$ O-orbitals
(dsp-model). The dominant 
parameters ($>0.1$eV)  for in-plane hopping are
given in the first rows of Table~\ref{andp}. Here $\br$ labels the O
on the rung of one ladder while $\bl$ labels either a Cu or an O on
the {\it upper} leg of the ladder. Vectors $\ve_x=(a/6,0,0)$ and
$\ve_z=(0,0,c/2)$ gives the translation 
vector from one O(Cu) to the neighboring Cu(O) in the 
respective direction. The
notation is such that $t_{d(\bl)p_x(+\ve_x)}$ denotes the hopping between
a $3d$ Cu-orbital at $\bl$ on the upper leg to the
neighboring  $2p_x$ O-orbital at $\bl+\ve_x$. The on-site 
$t_{s(\bl)d(\bl)}$ hopping
is non-zero as
consequence of the downfolding of all other bands in the absence of local
four-fold symmetry.

The on-site energies of the rung-oxygen $\epsilon(\br)$ 
is slightly larger than that
 of the leg-oxygen $\epsilon(\bl)$
due to the  local environments.
The distance between Cu and O is $r_l=1.98$\AA\ $\| \hat{z}$ and
$r_r=1.92$\AA\  $ \| \hat{x}$ implying a larger hopping along 
the $x$-direction according to $t_{pd}\propto 1/r^{4}$ and explaining the
anisotropy for the $t_{pd}$'s listed in first row of Table \ref{andp}.
Moreover
hopping processes involving $s$-orbitals are large with
non-negligible $t_{sd}$ hoppings. They will contribute
to the $t_{pd}$ hopping in 
second and higher orders through paths like $d-s-p$, $d-s-s-p$,  etc.
By downfolding the $s$-orbitals, 
 these processes will strongly renormalize the
$t_{pd}$ such that 
$t_{d(\bl)p_z(+\ve_z)}>t_{d(\bl)p_x(-\ve_x)}$.
Lastly downfolding the p orbitals gives the effective (d)-model. Results are
given in  third column of Table~\ref{par}. They are consistent with our
previous results. The anisotropy between $t_{\perp,1}$ and $t_{\|,1}$ 
can now be seen to be a consequence of different $t_{pd}$ hopping
due to the downfolding of the $s$ bands. 

	The exchange interaction $J$ between spins are usually difficult to
estimate a priori. In perturbation theory, $J$ scales as $t_{pd}^4$ but given
the large value of $t_{pd}$ relative to $(\epsilon_{p}-\epsilon_{d})$ the 
results are not reliable. Our analysis shows that the energy difference
$(\epsilon_{p}-\epsilon_{d})$ does not differ much between rung and legs
but $t_{pd}$ does.
 A ratio 
$J_{\perp,1}/J_{\|,1}<1$ is expected consistently 
 with previous results\cite{Johnston,Troyerpc,eccleston}.
This will disfavor the hole
coupling and  the RVB liquid state and may explain why holes are unbound 
for $T>100$K, however this point will not be discussed further. 

	In the following, two limiting  estimates of
 the conductivity are investigated. 
	First, the band structure model 
limit is considered ignoring  magnetic interactions although they must be
important.  In a metallic ground state, the conductivity  reduces to an
integral over the Fermi surface.  
\begin{equation}
	\sigma_{ij} =	\frac{2e^2}{(2\pi)^2}\int d\bk \tau(\epsilon(\bk))
	v_i(\bk) v_j(\bk) \left( -\frac{\partial f}{\partial \epsilon} \right).
\end{equation}
	  Considering an electron filling of $n=0.8$
the ratio between the conductivity perpendicular and parallel to the
ladder can then be simply computed yielding very large values i.e.
\mbox{$\sigma_{\|}/\sigma_{\perp}\simeq  75, 90 \mathrm{,\; and\; }104$},
for the parameter in Table~\ref{par}. Thus
roughly one has  
\begin{equation}
	\sigma_{\|}/\sigma_{\perp}\simeq  100.
\end{equation}
	This  large anisotropy is a consequence of a warped Fermi-surface
with very small Fermi velocity perpendicular to the ladders.

	Second, as discussed previously, the resistivity data 
perpendicular to the ladder indicates 
a mean-free path smaller than the inter-ladder spacing. Holes seem to be
confined to ladders and to hop incoherently between them. A detail structural
refinement of Ca11 \cite{muromachi} shows a complex distortion pattern of the
CuO$_2$-chains leading to substantial potential variations between rungs due
to the proximity of apical  O-ions at $\sim 50 \%$ of the rungs
An estimate of the interladder conductivity can be made
by considering the limit where ladders
form quasi one-dimensional  
metallic systems weakly coupled to each other. The conductivity is thus
a consequence of the  inter-ladder hopping term 
\begin{equation}
	H=\sum_{\bk} T_{k} c^{\dag}_{1,k} c_{2,k} +\hc.
\end{equation}
	where $T_{\bk}=\sum_i h_{\perp,i}(k)$. 
 The conductance $\sigma_{\perp}$ is given by 
\begin{equation}
	\sigma_{\perp}=4\pi \frac{e^2}{\hbar} {\cal
N(\epsilon_F)}^2|T(k_F)|^2,  
\end{equation}
with $\cal N(\epsilon_F)$ denoting the density of state at $E_F$
in the quasi-one dimensional metallic system.
	Results for the different parameter sets are $\sigma_{\perp} \simeq
0.1,0.09,$ and $0.074$, respectively. They are thus all close the value of 
\begin{equation}
	\sigma_{\perp} \simeq   0.08 \;\mathrm{m\Omega^{-1} cm^{-1}},
\end{equation}
	which corresponds well with the experimental result at
\mbox{$T \gs 100$K}. 

	In this paper, estimates of the hopping matrix elements based on
 LDA calculations gave three main results. First, the effective
intra-ladder hopping between n.n.\  are not the same,
$t_{\|}\neq t_{\perp}$. This was
explained as the consequence of anisotropic $t_{pd}$ in the ($pd$)-model
due to effective hopping through paths involving Cu $s$-states.
 Second, inter-ladder hopping  is  much smaller than
intra-ladder  and longer range hopping must be included.
Third, estimates of the conductivity in the model where holes are
unbound and 
confined into the ladder give good agreement with the experiment at
temperatures {$T  \gs 100$K}.\\
\hspace*{3mm}One of us (T.~M.) thanks the ``Fond National Suisse'' for
financial support. 
We would like to thank O.~K.~Andersen and M.~Troyer for useful and fruitful
discussions.

%
%
\begin{table}
\[
	\begin{array}{|c|rr|rr|}\hline
	&\multicolumn{2}{c|}{\srqcuo} & \multicolumn{2}{c|}{\srcuo} \\
	\cline{2-5}
	& \multicolumn{1}{c}{\mathrm{b-band}} &
	 \multicolumn{1}{c|}{\mathrm{a-band}} 
	 & \multicolumn{1}{c}{\mathrm{b-band}} &
	 \multicolumn{1}{c|}{\mathrm{a-band}} \\\hline
	\epsilon_0 & -0.31 & 0.46   & -0.44 & 0.35 \\
	h_{\|,1}        & 0.41  & 0.59  &  0.45 & 0.68\\
	h_{\|,2}        & 0.08  & 0.07  &  0.08 & 0.07\\ 
	h_{\perp,1}& 0.07  & 0.03  & 0.07 & 0.03 \\
	h_{\perp,2}& 0.00  &-0.04  & 0.00 & -0.04\\\hline
	\end{array}
\]
\caption{The hopping parameters in eV for \srqcuo and \srcuo in the rung scheme.}
\label{tabArai}
\end{table}
\begin{table}
\[
	\begin{array}{|c|r|r|r@{,}|}\hline
	& & \multicolumn{2}{c|}{\srcuo} \\\cline{3-4}
	& \multicolumn{1}{r|}{\srqcuo}  & \mathrm{Fit} &
	\multicolumn{1}{r|}{\mathrm{Downfolding}} \\\hline 
	\epsilon_0  &   0.075  &- 0.045   & -2.476\\
	t_{\|,1}    &   0.500  &  0.565   &  0.537\\
	t_{\perp,1} &   0.385  &  0.395   &  0.351  \\
	t_{\perp,2} &   0.040  &  0.040   &  0.018  \\
	t_{\perp,3} &   0.050  &  0.050   &  0.050  \\
	t_{\perp,4} & - 0.090  &- 0.115   & -0.124 \\
	t_{\perp,5} &   0.040  &  0.040   &  0.053  \\
	t_{\|,2}    &   0.075  &  0.075   & 0.106 \\
	t_{\perp,6} &   0.005  &  0.005   & 0.012  \\
	t_{\perp,7} & - 0.020  &  -0.020  & -0.023\\\hline
	\end{array}
\]
\caption{The parameters in eV from direct fitting and downfolding of the LDA
bands.}
\label{par}
\end{table}
\end{multicols}
\onecolumn
\begin{table}[p]
	\[
	\begin{array}{|l|*{7}{c}|}\hline
	& \epsilon_d &\epsilon_{p(\br)} & \epsilon_{p(\bl)}
	& t_{d(\bl)p_z(+\ve_z)} & t_{d(\bl)p_x(-\ve_x)}&
	t_{d(\bl)p_x(+\ve_x)}
	&  t_{p_z(\bl)p_z(+\ve_x+\ve_z)} \\\hline  

	\mathrm{(dsp)} & -3.82 & -3.66 & -4.00 & 0.74 & 0.85 &
	-0.85 & 0.30 \\\hline  
	\mathrm{(dp)}   & -3.83   & -4.59 & -4.53 & 0.65  & 0.51
	& -0.63  &  0.22\\ \hline\hline 

	& \multicolumn{2}{c}{t_{p_x(\bl)p_x(+\ve_x+\ve_z)}} &
	\multicolumn{2}{c}{t_{p_z(\bl)p_x(+\ve_x+\ve_z)}} &  
	t_{p_z(\bl)p_x(-\ve_x+\ve_z)}	 & 
	t_{p_x(\bl)p_x(-\ve_x+\ve_z)} &
	t_{p_z(\bl)p_z(+2\ve_z)} \\\hline
	\mathrm{(dsp)}&
	\multicolumn{2}{c}{0.30} & 
	\multicolumn{2}{c}{0.01} & -0.13 & 0.28 & -0.42 \\\hline
	\mathrm{(dp)} &  
	\multicolumn{2}{c}{ 0.62} &
	\multicolumn{2}{c}{ 0.62} & -0.64 &  0.19 & -0.05 \\\hline\hline

	& \multicolumn{2}{c}{t_{p_x(\bl)p_x(+2\ve_z)}} &
	\multicolumn{2}{c}{t_{p_x(\br)p_x(+2\ve_x)}}&
	 \epsilon_s & 	t_{s(\bl)d(\bl) } & 
	 t_{s(\bl)d(+2\ve_z) } \\\hline
	\mathrm{(dsp)}& \multicolumn{2}{c}{-0. 17} &
	\multicolumn{2}{c}{-0.62 } &  2.43   &  -0.24 & -0.14\\\hline
	\mathrm{(dp)} &\multicolumn{2}{c}{-0.09} &
	 \multicolumn{2}{c}{+0.31} & \mbox{---} & \mbox{---}  & \mbox{---}
	\\\hline\hline

	& \multicolumn{2}{c}{t_{s(\bl)s(+2\ve_z)}} &
	 \multicolumn{2}{c}{t_{s(\bl)d(-2\ve_x) }} & 
	 t_{s(\bl)s(-2\ve_x) } & t_{s(\bl)p_z(+\ve_z)}  
	& t_{s(\bl)p_x(\pm \ve_x) }  \\\hline
	\mathrm{(dsp)}& \multicolumn{2}{c}{-0.32} &
	\multicolumn{2}{c}{0.20} & -0.44 & 1.89 & \pm2.10 
	\\\hline\hline
	\end{array}\]
\caption{Hopping parameters in eV from the downfolding method. Notation is
explained in the text. \label{andp}}
\end{table}


\begin{references}
\bibitem{newkx}
We use the notation \mbox{$k_x a/2 \longrightarrow k_x$} and \mbox{$k_zc
\longrightarrow k_z$}.
\end{references}
\end{document}